\documentclass[12pt]{article}
\usepackage{graphics}
\usepackage{epsfig}
\begin{document}
\newcommand{\beq}{\begin{equation}}
\newcommand{\eeq}{\end{equation}}
\newcommand{\bea}{\begin{eqnarray}}
\newcommand{\eea}{\end{eqnarray}}
\newcommand{\eps}{\varepsilon}
\newcommand{\Fs}{\mbox{\scriptsize F}}

{\bf The role of the boundary conditions in the Wigner-Seitz
approximation applied to the neutron star inner crust.}

\centerline{ M.~Baldo$^{a}$, E.E.~Saperstein$^{b}$ and
S.V.~Tolokonnikov$^{b}$ }

\vskip 0.5 cm \centerline{$^a$INFN, Sezione di Catania, 64 Via
S.-Sofia, I-95123 Catania, Italy} \centerline {$^{b}$ Kurchatov
Institute, 123182, Moscow, Russia }
\vskip 0.5 cm

\begin{abstract}

The influence of the boundary conditions used in the Wigner-Seitz
approximation applied to the neutron star inner crust is examined.
The generalized energy functional method which includes neutron
and proton pairing correlations is used. Predictions of two
versions of the boundary conditions are compared with each other.
The uncertainties in the equilibrium configuration ($Z,R_c$) of
the crust, where $Z$ is the proton charge and $R_c$ the radius of
the Wigner-Seitz cell, correspond to variation of $Z$ by 2 -- 6
units and of $R_c$, by 1 -- 2 fm. The effect of the boundary
conditions is enhanced at increasing density. These uncertainties
are smaller than the variation of $Z$ and $R_c$ coming from
the inclusion of pairing. The value of the pairing gap itself, especially
at high density, can depend on the boundary condition used.
\end{abstract}
\vskip 0.2 cm
\noindent
PACS : 26.60.+c,97.60.Jd,21.65.+f,21.60.-n,21.30.Fe
\vskip 0.4 cm
\par
In the last two decades the interest on the structure of the
neutron star inner crust has been stimulated by the increasing
number of observational data on the pulsar glitches. The latter
are commonly explained in terms of the dynamics of superfluid
vortices within the inner crust of neutron stars (see \cite{Pet}
and Refs. therein). By ``inner crust'' one usually indicates the
part of the shell of a neutron star with sub-nuclear densities
$0.001 \rho_0 \le \rho \le 0.5 \rho_0 $, where $\rho_0$ is the
normal nuclear density.
 According to present-day ideas, the bulk
of the inner crust consists mainly of spherically symmetrical
nuclear-like clusters which form a crystal matrix immersed in a
sea of neutrons and virtually uniform sea of electrons.
 Such a picture was first justified microscopically in the classical
paper by Negele and Vautherin (NV) \cite{NV} within the
Wigner-Seitz (WS) approximation. Up to now, the WS method remains
to be quite popular in this field. Only recently a more consistent
band theory was developed for the deep (high
density) layers \cite{CCH}, where the ``lasagna'' or ``spaguetti''
structure of the crust matter is supposed to be favored, and for
the outer (low density) layers as well \cite{NC}. As far as the
band theory is quite complicated, the WS method is usually
considered as the most practical one for systematic investigation
of the inner crust structure in the whole density interval.

 NV used for describing the matter of a neutron star crust
a version of the energy functional method with  density dependent
effective mass $m^*(\rho)$. In fact, it is very close to the
Hartree-Fock method with effective Skyrme forces. For a fixed
average nuclear density $\rho$, the nuclear (plus electron) energy
functional is minimized for the spherical WS cell of the radius
$R_c$. A cell contains $Z$ protons (and electrons) and $N=A-Z$
neutrons ($A=(4\pi/3)R_c^3\rho$). In addition, the
$\beta$-stability condition,

\beq \mu_n-(\mu_p+\mu_e)=0, \label{mu} \eeq has to be fulfilled,
where $\mu_n$, $\mu_p$ and $\mu_e$ are the chemical potentials of
neutrons, protons and electrons, respectively. The minimization
procedure is carried out for different values of $Z$ and $R_c$.
The equilibrium configuration ($Z,R_c$) at the considered density
corresponds to the absolute minimum in energy among all these
possible configurations.

Application of the variational principle to the NV energy
functional for a WS cell results in the set of the
Shr\"odinger-type equations for the single particle neutron
functions $\phi_{\lambda}({\bf r})=R_{nlj}(r) \Phi_{ljm}({\bf
n})$, with the standard notation. The radial functions
$R_{nlj}(r)$ obey the boundary condition (BC) at the point
$r=R_c$. There exist different kinds of the BC. NV used the
following one:

 \beq R_{nlj}(r=R_c)=0
 \label{bco}
\eeq for odd $l$, and \beq \left(\frac
{dR_{nlj}}{dr}\right)_{r=R_c}=0, \label{bce} \eeq
 for even ones.
Let us denote it as BC1. The use of this BC has been partly justified
by physical considerations in NV, but the dependence of the results on
the BC has never been discussed in detail. It is the purpose of the paper
to study this problem at a quantitative level and to establish the
corresponding uncertainity, which is inherent to the WS method applied
to neutron star crust.
%In this paper, we examine the dependence
%of the results of applying the WS method to the neutron star inner
%crust on the kind of the BC.
For this aim, we compare results
obtained for the BC1 with those found for an alternative kind of
the BC (BC2) when Eq.~(\ref{bco}) is valid for even $l$ whereas
Eq.~(\ref{bce}), for odd ones. In principle, two additional kinds
of the BC exist when Eq.~(\ref{bco}) or Eq.~(\ref{bce}) is used
for any $l$. As it was noted by NV, these BC have an obvious
drawback for the case of the neutron star inner crust as far as
they lead to an unphysical irregular behavior of the neutron
density $\rho_n(r)$ in vicinity of the point $r=R_c$. Indeed,
$\rho_n(r)$ vanishes in this point in the first case and has a
maximum in the second one. On the contrary, $\rho_n(r)$ is almost
constant nearby the point $r=R_c$ in the case of the BC1 or BC2.

It should be noted that the pairing effects were not taken into
account in \cite{NV} since  it was supposed that they are not
important for the structure of the crust. The reason of such an
assumption is the rather small contribution of the pairing effects to the
total binding energy of the system under consideration. Recently,
we have generalized the NV approach to describe the inner crust by
explicitly including the neutron and proton pairing correlations
\cite{crust1,crust2,crust3,crust4} in a self-consistent way. It
turned out that in the whole interval of $\rho$ the equilibrium
configuration ($Z,R_c$) changes significantly due to pairing.

We used the generalized energy functional method  \cite{Fay} which
incorporates the pairing effects into the original Kohn-Sham (KS)
\cite{KS} method. In this approach, the interaction part of the
generalized energy functional (GEF) depends, on equal footing, on
the normal densities $\rho_n , \rho_p$, and the abnormal ones,
$\nu_n, \nu_p$, as well: \beq E_{int} = \int d {\bf r} {\cal
E}_{int}(\rho(\bf r ),\nu({\bf r})),\label{GEF} \eeq where ${\cal
E}_{int}$ is the GEF density. It is the sum of two components, the
normal and the anomalous (superfluid) ones: \beq {\cal E}_{int} =
{\cal E}_{norm}(\rho_{\tau})+ {\cal E}_{an}(\rho_{\tau},
\nu_{\tau}), \label{Func} \eeq where $\tau=n,p$ is the isotopic
index. Just as in the KS method, the prescription $m^*=m$ holds to
be true.

To describe the central part of a WS cell with the nuclear cluster
inside we used the phenomenological nuclear GEF ${\cal E}^{ph}$ by
Fayans et al. \cite{Fay} which describes properties of the
terrestrial atomic nuclei with high accuracy. For describing
neutron matter surrounding the cluster we  used a microscopic
energy functional ${\cal E}^{mi}$ for neutron matter based on the
Argonne NN potential  $v18$ \cite{v18}. The ansatz of
\cite{crust3,crust4} for the complete energy functional is  a
smooth matching of the phenomenological and the microscopic
functionals at the cluster surface:
 \beq
{\cal E}(\rho_{\tau}({\bf r}),\nu_{\tau}({\bf r})) = {\cal
E}^{ph}(\rho_{\tau}({\bf r}),\nu_{\tau}({\bf r})) F_m( r)+ {\cal
E}^{mi}(\rho_{\tau}({\bf r}),\nu_{\tau}({\bf r}))(1 - F_m(r)),
\label{tot} \eeq where the matching function $F_m(r)$ is a
two-parameter Fermi function: \beq
F_m(r)=(1+\exp((r-R_m)/d_m))^{-1}. \label{match} \eeq

Eq. (\ref{tot}) is applied  both to the normal and to the
anomalous components of the energy functional. After a detailed
analysis, the matching parameters were chosen as follows. The
diffuseness parameter was taken to be equal to $d_m{=}0.3\;$fm for
any value of the average baryon density of the inner crust and for
any configuration ($Z,R_c$). As to the matching radius $R_m$, it
should be chosen anew in any new case, in such a way that the
equality \beq \rho_p(R_m)=0.1 \rho_p(0) \label{match1} \eeq holds.
In this case, on one hand, for $r <\; R_m$ neutrons and protons
coexist inside the nuclear-type cluster, and the use of a
realistic phenomenological energy functional seems reasonable. On
the other hand, at $r >\; R_m$ one can neglect the exponentially
decaying proton "tails" and consider the system as a pure neutron
matter for which an adequate energy functional microscopically
calculated can be used. The same matching parameters  were used
for normal and anomalous parts of (\ref{tot}). As far as
practically all the protons are located inside the radius $R_m$,
the matching procedure concerns, in fact, only neutrons, protons
being described with the pure phenomenological nuclear GEF.

It is worth to mention that for neutron matter region, the ansatz
is, in fact, the LDA for the microscopic part of the GEF. As it is
commonly known, the LDA works well only provided the density is
smoothly varying, whereas it fails in the surface region with a
sharp density gradient.  The above choice of the matching
procedure and the values of the parameters guarantees that this
region of a sharp density variation  is mainly governed by the
phenomenological nuclear part of the GEF which "knows how to deal
with it".

For the microscopic part of the normal component of the total
energy functional (\ref{tot}) we follow refs. \cite{crust3,crust4}
and take the EOS of neutron matter calculated in \cite{B-V} with
the Argonne $v18$ potential on the basis of Brueckner theory,
taking into account a small admixture of 3-body force. Its
explicit form could be found in the cited articles. The
microscopic part of the anomalous component of the GEF in
\cite{crust3,crust4} was calculated for the same $v18$ potential
within the BCS approximation.

As calculations of \cite{crust1,crust2,crust3,crust4} have shown,
the pairing correlations influence the equilibrium values
($Z,R_c$) significantly. To explain this effect, it is instructive
to analyze the $\beta$-stability condition (\ref{mu}). As far as
electrons in the inner crust of a neutron star are
ultra-relativistic, the following relation is valid: $\mu_e \simeq
(9\pi Z/4)^{1/3}/R_c$. By substituting it into Eq.~(\ref{mu}), one
finds \beq Z \simeq \frac {4} {9 \pi} (\mu_n-\mu_p)^3 R_c^3.
\label{Zc} \eeq  The influence of pairing on the chemical
potentials $\mu_n$ and $\mu_p$ is much stronger than that on the
total binding energy. Their variation may be of the order of the
gap value $\Delta \simeq$1--2$\;$MeV. Such a variation of $\mu_n$
or $\mu_p$ may lead to a sizable change of the equilibrium value
of $Z$ as far as the difference ($\mu_n - \mu_p$) is raised to the
third power in Eq.~(\ref{Zc}). The estimate of the change of $Z$
induced by this variation is as follows: $\delta
Z{=}3Z\delta(\mu_n-\mu_p)/ (\mu_n-\mu_p)$. For average values of
$k_{\Fs}$, the difference $\mu_n-\mu_p \simeq 50 \div 70\;$MeV,
hence $\delta Z$ could reach several units of $Z$. An additional
change of the $Z$ value may appear due to a variation of $R_c$.

Besides, as it can be seen in Fig. 1, the binding energy $E_B$ is
rather flat function of $Z$ and
 different local minima $E_B(Z)$ have often close values of $E_B$.
Therefore their relative position may change after switching off
the pairing since in general the corresponding contribution to
$E_B$ is an irregular function of $Z$. Such a situation does often
occur within the WS approach, especially for high density values,
due to the shell-type effect in the single-particle neutron
spectrum. An example is discussed below.

\begin{figure}
\includegraphics [height=160mm,width=100mm]{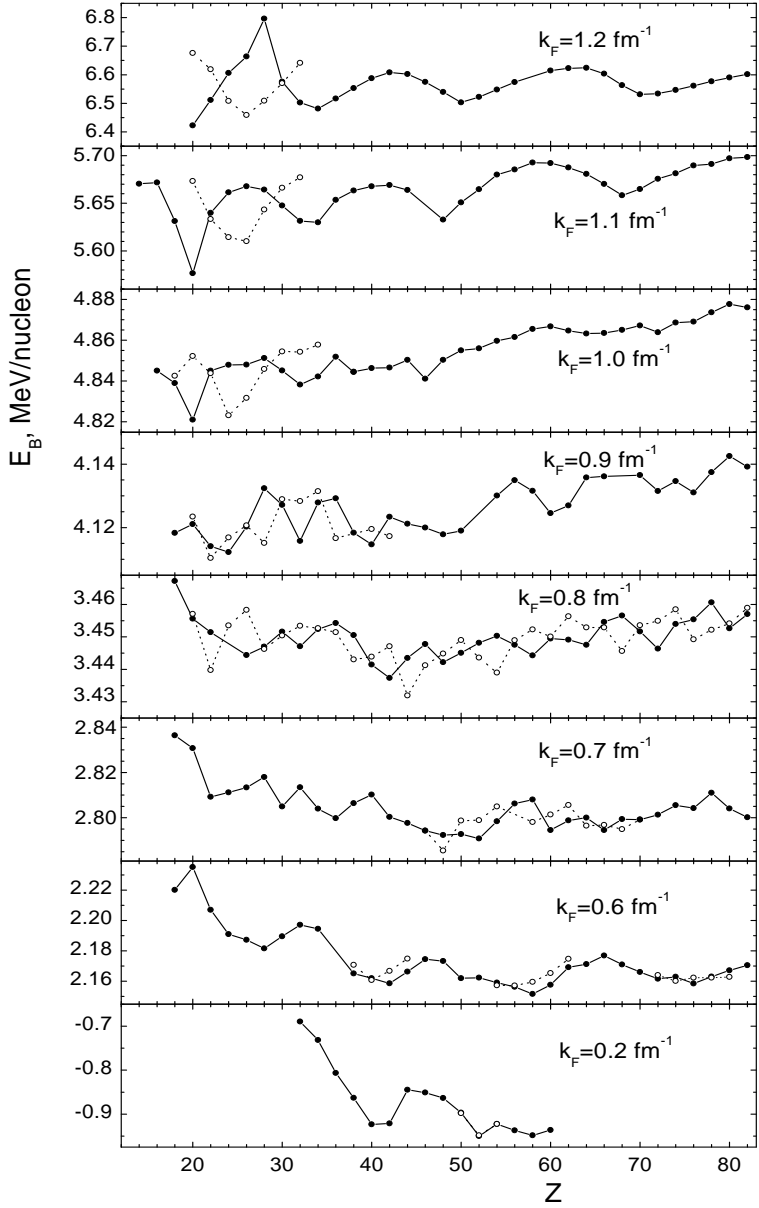}
\caption{ Binding energy per nucleon for various $k_{\Fs}$ in the
case of the BC1 (solid circles connected with the solid lines) and
the BC2 (open  circles connected with the dotted lines).}
\end{figure}

 In the calculations of \cite{crust1,crust2,crust3,crust4}
the NV boundary condition, BC1, was used. Here we repeat the
analysis for the case of the boundary condition BC2. Results for
the binding energy are shown in Fig. 1. Just as in
\cite{crust3,crust4} only even values of Z are used. The detailed
comparison is made for $k_{\Fs}{=}0.8\;$fm$^{-1}$. Although the
two curves $E_B(Z)$ are quite different, the positions of local
minima for the BC1 and BC2 are close to each other, the distance
being equal to 2 or 4 units of Z. What is of primary importance,
the relative position of local minima for BC2 is the same as for
BC1. In particular, the positions of the absolute minimum almost
coincide ($Z{=}52$ for BC1 and $Z{=}54$ for BC2). These
observations permit us to simplify calculations for other values
of $k_{\Fs}$. In the case of the BC2, we limit ourselves mainly
with the analysis of a vicinity of the absolute minimum for the
BC1. The neighborhood of other local minima was analyzed only in
the case if they have values of $E_B(Z)$
 close to that corresponding to the absolute minimum. It
turned out that there is no value of $k_{\Fs}$  for which the
relative position of a local minimum and of the absolute one for
the BC1 and BC2 is different. In addition to systematic
calculations for $k_{\Fs}{=}0.6 \div 1.2\;$fm$^{-1}$, we made an
extra one for a small density, $k_{\Fs}{=}0.2\;$fm$^{-1}$, in
vicinity of the neutron drip point. In the last case, two curves
corresponding to BC1 and BC2 practically coincide. For all other
values of $k_{\Fs}$ the absolute minima are shifted by 2, 4 or
even 6 units of $Z$.

\begin{table}[h!]
\caption{Comparison of properties of equilibrium configurations of
the WS cell for two different kinds of the boundary condition}
\bigskip

\begin{tabular}{|c|c|c|c|c|c|c|c|c|c|}
\hline
  \raisebox{-6pt}{$k_{\rm F}$} & \raisebox{-6pt}{$Z$} &
  \multicolumn{2}{|c|}{$R_{\rm c},\;$fm}\rule{0pt}{14pt}&
  \multicolumn{2}{|c|}{$E_B,\;$MeV}&
  \multicolumn{2}{|c|}{$\mu_{\rm n},\;$MeV}&
  \multicolumn{2}{|c|}{$\Delta_{\rm F},\;$MeV}\\
\cline{3-10}
 \rule{0pt}{13pt} && BC1 & BC2 & BC1 & BC2 & BC1 & BC2 & BC1 &  BC2\\
\hline
 0.2  & 52 & 57.18 & 57.10 &-0.9501 &-0.9483 & 0.1928 & 0.1942 & 0.04 & 0.05\\
\hline
  0.6 & 58 & 37.51 & 37.48 & 2.1516 & 2.1596 & 3.2074 & 3.2226 & 1.92 & 1.89\\
      & 56 & 36.97 & 36.95 & 2.1563 & 2.1572 & 3.2173 & 3.2193 & 1.91 & 1.89\\
\hline
  0.7 & 52 & 32.02 & 32.04 & 2.7908 & 2.7989 & 3.9876 & 4.0107 & 2.30 & 2.25\\
      & 48 & 31.16 & 31.14 & 2.7924 & 2.7856 & 4.0069 & 3.9873 & 2.29 & 2.32\\
\hline
  0.8 & 42 & 26.90 & 26.91 & 3.4373 & 3.4471 & 4.8454 & 4.8561 & 2.56 & 2.45\\
      & 44 & 27.29 & 27.30 & 3.4435 & 3.4319 & 4.8553 & 4.8198 & 2.53 & 2.56\\
\hline
  0.9 & 24 & 20.26 & 20.30 & 4.1123 & 4.1169 & 5.7340 & 5.7986 & 2.64 & 2.51\\
     &  22 & 19.87 & 19.70 & 4.1141 & 4.1104 & 5.7861 & 5.7170 & 2.62 & 2.54\\
\hline
  1.0 & 20 & 16.69 & 16.90 & 4.8210 & 4.8522 & 6.8525 & 6.7424 & 2.02 & 2.52\\
      & 24 & 18.29 & 18.22 & 4.8479 & 4.8231 & 6.8446 & 6.8920 & 2.52 & 2.29\\
\hline
  1.1 & 20 & 14.99 & 15.33 & 5.5765 & 5.6733 & 7.4288 & 8.0446 & 1.32 & 2.32\\
      & 26 & 16.75 & 17.08 & 5.6677 & 5.6100 & 7.9680 & 8.5398 & 2.28 & 2.02\\
\hline
  1.2 & 20 & 13.68 & 13.95 & 6.4225 & 6.6762 & 8.5814 & 9.1898 & 1.21 & 1.56\\
      & 26 & 15.21 & 14.89 & 6.6639 & 6.4587 & 9.0825 & 9.3413 & 1.25 &
      0.86\\[1mm]
\hline
\end{tabular}
\end{table}

Comparison of different properties of the equilibrium
configuration of the WS cell for various values of $k_{\Fs}$ in
the case of the BC1 and BC2 is presented in  Table 1.  There are
two lines for every value of $k_{\Fs}$. The first one is given for
the $Z$ value corresponding to the minimum of $E_B$ in the case of
the BC1, the second one, for the BC2. The only exception is
$k_{\Fs}{=}0.2\;$fm$^{-1}$ when these two values of $Z$ coincide.
In the last two columns, the average value $\Delta_{\Fs}$ of the
diagonal matrix element of the neutron gap at the Fermi surface is
given. The averaging procedure involves 10 levels above $\mu_n$
and 10 levels below. One can see that the influence of the BC is
enhanced at increasing values of $k_{\Fs}$. Especially strong
variation of $\Delta_{\Fs}$ and $\mu_n$ takes place in the cases
of $k_{\Fs}{=}1.1\;$fm$^{-1}$ and $k_{\Fs}{=}1.2\;$fm$^{-1}$ (Fig.
2).
\begin{figure}[h!]
\vspace{2mm}
\includegraphics [height=100mm,width=120mm]{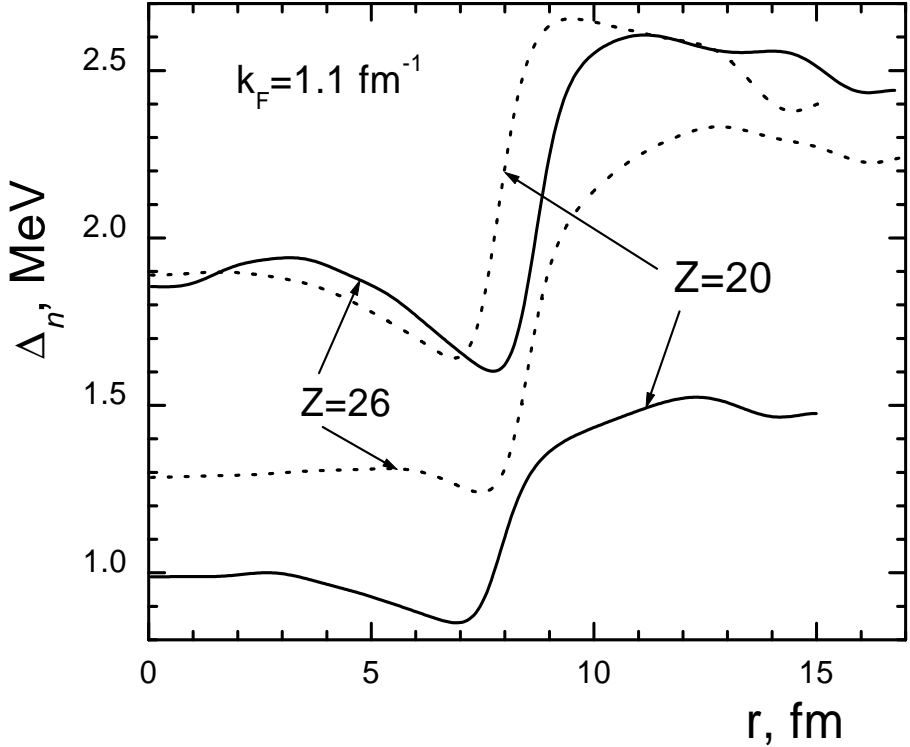}%
 \caption{ The neutron gap for
$k_{\Fs}{=}1.1\;$fm$^{-1}$, $Z{=}20$ and $Z{=}26$, in the case of
the BC1 (solid lines) and the BC2 (dashed lines).}
\end{figure}
To illustrate the influence of the BC to the neutron gap in the
first case, the gap function $\Delta_n(r)$ is drawn for both
values of $Z$ and both kinds of the BC. The most strong variation
of the gap occurs in the case of $Z{=}20$. To understand  the
reason of such strong effect, we draw the neutron single particle
spectrum $\eps_{\lambda}$ for this value of $Z$ in Fig. 3 for the
BC1 (the left half of the figure) and the BC2 (the right one). The
position of the chemical potential $\mu_n$ is shown with dots. The
two spectra are absolutely different. The reason is the shift
$\Delta \eps_{\lambda}$  of each $\lambda$-level going from BC1 to
BC2. The value of this shift is approximately equal to a half of
the distance between two neighboring levels with the same ($l,j$),
 the sign of the shift being opposite for even and odd $l$. The
absolute value of the shift is proportional to $1/R_c^2$ and grows
at increasing values of $k_{\Fs}$. The corresponding shifts are
shown in Fig. 3 for two states, 2$j_{13/2}$ and 1$n_{23/2}$, which
are the neighbors of $\mu_n$ in the BC1 case. On average, the
spectrum is quite dense, however in both cases there is a shell
type structure with rather wide intervals between some neighboring
levels. If one deals with a big inter-level space in vicinity of
$\mu_n$, as in the BC1 case in Fig. 3, one usually obtains a dense
set of levels in this region when going to the opposite kind of
the BC. In the case of the BC2, big intervals are far from the
Fermi surface and do not influence significantly the value of the
neutron gap. On the contrary, in the case of the BC1 $\mu_n$  is
situated just inside such an interval that suppresses the gap
significantly. In principle, the neutron gap could vanish if the
interval was wider.

\begin{figure}[t!]
\centerline{\includegraphics [height=110mm,width=80mm]{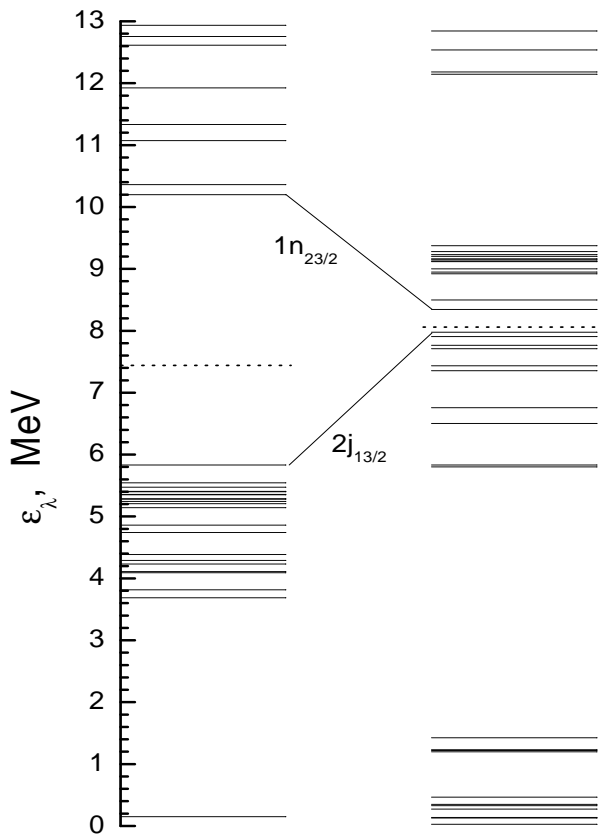}}
 \caption{ The neutron single-particle spectrum
$\eps_{\lambda}$ for $k_{\Fs}{=}1.1\;$fm$^{-1}$, $Z{=}20$, in the
case of the BC1 (left) and the BC2 (right).}
\end{figure}

As we see,  there are internal uncertainties inherent to the WS
 method applied to the neutron star inner crust which
originate from the kind of the BC used.
 Only in the case of very small density nearby the neutron drip
 point  predictions of the BC1 and BC2 versions are practically
 identical. If we deal with $k_{\Fs} \ge 0.6\;$fm$^{-1}$,
 the  uncertainty in the equilibrium value of $Z$ is between 2 and
 6 units, growing with increase of $k_{\Fs}$. The
 uncertainty in the value of $R_c$ is, as a rule, about 1 fm and
 only for $k_{\Fs}{=}1.1\;$fm$^{-1}$ it turns out to be about 2
 fm. However, the value of these uncertainties is less than the variation
 of the equilibrium configuration ($Z,R_c$) connected with
 the pairing effects \cite{crust4}.
 In the case of high densities, $k_{\Fs} \ge 1\;$fm$^{-1}$, the
 most important uncertainty occurs in the value of the neutron gap
 $\Delta_n$. It originates from the shell effect in the
 neutron single-particle spectrum which is rather pronounced in
 the case of big $k_{\Fs}$ and, correspondingly, small $R_c$ values.

\vskip 0.5 cm The authors thank  N.E. Zein for valuable
discussions. This research was partially supported by the Grant
NS-1885.2003.2 of The Russian Ministry for Science and Education.

{}


\begin{thebibliography}{0}


\bibitem{Pet} C.J. Pethick, D.G. Ravenhall, Ann. Rev. Nucl. Part.
 Sci.  45 (1995) 429.

\bibitem{NV} J. Negele, D. Vautherin, Nucl. Phys. A 207 (1973) 298.

\bibitem{CCH} B. Carter, N. Chamel, P. Haensel, Nucl. Phys.
 A 748 (2005) 675.

\bibitem{NC}  N. Chamel, arXiv preprint: nucl-th/0512034

\bibitem{crust1}  M. Baldo, U. Lombardo, E.E. Saperstein, S.V. Tolokonnikov,
JETP Lett.  80 (2004) 595.

\bibitem{crust2}  M. Baldo, E.E. Saperstein, S.T. Tolokonnikov,
               Nucl. Phys. A 749 (2005) 42.

\bibitem{crust3}  M. Baldo, U. Lombardo, E.E. Saperstein, S.V. Tolokonnikov,
Nuc. Phys. A 750 (2005) 409.

\bibitem{crust4}  M. Baldo, U. Lombardo, E.E. Saperstein, S.V. Tolokonnikov,
Phys. At. Nucl. 68 (2005) 1812.

\bibitem{Fay}  S.A. Fayans, S.V. Tolokonnikov, E.L. Trykov,  D. Zawischa,
Nucl. Phys. A 676 (2000) 49.

\bibitem{KS} W. Kohn, L.J. Sham, Phys. Rev. A 140 (1965) 1133.

\bibitem{v18}  R.B. Wiringa, V.G.J. Stoks, R. Schiavilla, Phys. Rev. C 51 (1995) 38.

\bibitem{B-V} M. Baldo, C. Maieron, P. Schuck, X. Vinas,
Nucl. Phys. A  736 (2004) 241.

\end{thebibliography}
\end{document}